\documentstyle[11pt]{article}

\def\be{\begin{equation}}
\def\ee{\end{equation}}
\def\bea{\begin{eqnarray}}
\def\eea{\end{eqnarray}}
\def\ba{\begin{array}}
\def\ea{\end{array}}
\def\bma{\left(\begin{array}}
\def\ema{\end{array}\right)}

\begin{document}

\title{ The upper triangular solutions to the three-state
\\ constant quantum Yang-Baxter equation }

\author{Jarmo Hietarinta\\
Department of Physics\\
University of Turku, 20500 Turku, Finland}

\date{June 16, 1993}

\maketitle

\begin{abstract}
In this article we present all nonsingular upper triangular solutions
to the constant quantum Yang-Baxter equation
$R_{j_1j_2}^{k_1k_2}R_{k_1j_3}^{l_1k_3}R_{k_2k_3}^{l_2l_3}=
R_{j_2j_3}^{k_2k_3}R_{j_1k_3}^{k_1l_3}R_{k_1k_2}^{l_1l_2}$ in three
state case. The upper triangular ansatz implies 729 equations for 45
variables. Fortunately many of the equations turned out to be simple
allowing us to start breaking the problem into smaller ones. In the
end we had a total of 552 solutions, but many of them were either
inherited from two-state solutions or subcases of others. The final list
contains 35 nontrivial solutions, most of them new.
\end{abstract}

\baselineskip=18pt

\section{Introduction}
In this paper we continue our work on systematically solving the
constant quantum Yang-Baxter equation
\be
R_{j_1j_2}^{k_1k_2}R_{k_1j_3}^{l_1k_3}R_{k_2k_3}^{l_2l_3}=
R_{j_2j_3}^{k_2k_3}R_{j_1k_3}^{k_1l_3}R_{k_1k_2}^{l_1l_2},
\label{YBE}
\ee
In our previous paper \cite{jh1} (for more details, see \cite{jh2}) we
solved completely the two-state problem, which involved 64
equations in 16 variables.

In general~(\ref{YBE}) contains $N^6$ equations for $N^4$ unknowns, so
in the present case when $N=3$ we have 729 equations for 81 variables.
At the moment this is too complicated for an exhaustive study and one
must proceed step by step with suitable ansatze.  The present ansatz
is based on the two-state results: it turned out \cite{jh1,hl1,hl2}
that all two-state nonsingular solutions were either
\begin{itemize}
\item Upper triangular: $R_{ij}^{kl}=0$, if $j>l$ or $j=l,\,i>k$, or
\item Even weight: $R_{ij}^{kl}=0$, if $k+l\ne i+j \pmod 2$.
\end{itemize}
(In thinking $R_{ij}^{kl}$ as an $N^2\times N^2$ matrix
we use the convention that the right-hand indices define the block.)
Thus upper triangularity turns out to be a good ansatz for finding
two-state solutions, and we hope it is equally productive when the
number of states is three.

\section{Simple solutions}
\label{simple}
Let us first of all note that the YBE does have some `easy' classes of
solutions. That is, the solutions are easy to construct from previous
solutions or for some other reason. [This does not mean that the other
related structures (e.g.\ algebras) are necessarily less interesting.]
In this section we will describe two such classes and solutions
belonging to them are not mentioned again when we discuss the results
of our search.

\subsection{Solutions inherited from lower number of states}
A solution to (\ref{YBE}) with a lower number of states can always be
dressed to become a higher-state solution. There are at least two ways
to do this.

\subsubsection{Diagonal dressing}
Let $\tilde R$ be an $M$-state solution of (\ref{YBE}) and $N>M$. Let
$\mu$ be a selection of $M$ numbers from $\{1,\dots,N\}$, and define the
$N$-state $R$-matrix as follows:
\be
R_{ij}^{kl}=\left\{\ba{ll}
\tilde R_{ij}^{kl}, & \mbox{ when }i,j,k,l\in\mu,\\
s_{ij}\delta_i^k\delta_j^l, & \mbox{ otherwise.}
\ea\right.\label{NDR}
\ee
When this is substituted into (\ref{YBE}) we find further conditions
on $s$ as follows:
\bea
\tilde R_{ij}^{kl}\,(s_{mi}s_{mj}-s_{mk}s_{ml})=0,\nonumber \\
\tilde R_{ij}^{kl}\,(s_{im}s_{ml}-s_{km}s_{mj})=0,\label{Dcon}\\
\tilde R_{ij}^{kl}\,(s_{im}s_{jm}-s_{km}s_{lm})=0,\nonumber
\eea
$\forall\, i,j,k,l\in\mu,\,m\not\in\mu$, no sum.  [It is interesting
to note that the conditions are exactly like those obtained in a
different context in \cite{Kempf}.]  A trivial solution to
(\ref{Dcon}) is given by $s_{ij}=1$, but there can also be others
(depending on the form of $\tilde R$).

In the present case we have $M=2$ and $N=3$ and there is only one
index outside $\mu$. As an example let us take $\mu=\{1,3\}$ and
\be
\tilde R=\bma{cc|cc}
1 & . & . & . \\
. & p & 1-pq & . \\
\hline
. & . & q & . \\
. & . & . & 1 \ema,\label{D2S1}
\ee
(for matrices we always write `.' in place of a `0', for better
readability) and then the only remaining condition from (\ref{Dcon}) is
$s_{12}s_{21}=s_{23}s_{32}$. Thus we can extend (\ref{D2S1}) to a
three-sate solution  with the four additional parameters $a,\,b,\,x,\,y$:
\be
R=\bma{ccc|ccc|ccc}
1 & . & . & . & . & . & . & . & . \\
. & a & . & . & . & . & . & . & . \\
. & . & p & . & . & . & 1-pq & . & . \\
\hline
. & . & . & by & . & . & . & . & . \\
. & . & . & . & x & . & . & . & . \\
. & . & . & . & . & b & . & . & . \\
\hline
. & . & . & . & . & . & q & . & . \\
. & . & . & . & . & . & . & ay & . \\
. & . & . & . & . & . & . & . & 1
\ema.
\ee

\subsubsection{Block dressing}
The starting point is as above but the new higher-sate solution is
constructed as follows:
\be
R_{ij}^{kl}=\left\{\ba{ll}
\tilde R_{ij}^{kl}, & \mbox{ when }i,j,k,l\in\mu,\\
\delta_i^kF_j^l, & \mbox{ when }j,l\in\mu,\,i,k\not\in\mu\\
G_i^k\delta_j^l, & \mbox{ when }i,k\in\mu,\,j,l\not\in\mu\\
\delta_i^k\delta_j^l, & \mbox{ otherwise.}
\ea\right.\label{NBR}
\ee
This ansatz leads to the conditions
\bea
(F\otimes F)\tilde R &=&\tilde R (F\otimes F),\nonumber\\
(1\otimes F)\tilde R(G\otimes 1)&=&(G\otimes 1)\tilde R(1\otimes F),
\label{ENBRC}\\
\tilde R(G\otimes G)&=&(G\otimes G)\tilde R,\nonumber\\
\left[ F , G \right] & = & 0 \nonumber.
\eea
The most important case of this is when $G=F^{-1}$, in which case the first
equation of (\ref{ENBRC}) is sufficient. An example of such an solution is
\be
R:=
\bma{ccc|ccc|ccc}
1 & . & 1 & . & . & . & -1 & . & c_9 \\
. & -1 & . & . & . & . & . & d_6 & . \\
. & . & 1 & . & . & . & . & . & c_9\\
\hline
. & . & . & g_6 & . & d_6g_6 & . & . & .\\
. & . & . & . & f_5 & . & . & . & .\\
. & . & . & . & . & g_6 & . & . & .\\
\hline
. & . & . & . & . & . & 1 & . &  - c_9\\
. & . & . & . & . & . & . & -1 & .\\
. & . & . & . & . & . & . & . & 1\ema,
\ee
which decomposes as
\be
\tilde R:=
\bma{cc|cc}
1 & 1 & -1 & c_9\\. & 1 & . & c_9\\
\hline
. & . & 1 &  - c_9\\. & . & . & 1\ema,
\quad
G:=
\bma{cc}
-1 & d_6\\. & -1\ema,
\quad
G:=
\bma{cc}
g_6 & d_6g_6\\. & g_6\ema.
\ee

\subsection{Solutions generated by commuting matrices}
The following solutions can be said to be inherited from lower dimension,
when dimension means the number of pairs of
indices. (This hierarchial structure is more obvious when one considers
extensions, see e.g. \cite{Zam:mn}.)

The result is simply the following: Let
$\{N(\alpha),M(\alpha)|\alpha\in I\}$ be a set of commuting $N\times N$
matrices, then it is easy to show that
\be
R_{ij}^{kl}=\sum_{\alpha\in I}N(\alpha)_i^k M(\alpha)_j^l
\ee
is an $N$-sate solution of (\ref{YBE}).

A commuting set of matrices can be simultaneously brought to the
Jordan canonical form, but the Jordan form does not have to be
diagonal. A two-state example is provided by
\be
R=\bma{cc|cc}
1 & a & b & c \\
. & 1 & . & b \\
\hline
. & . & 1 & a \\
. & . & . & 1 \ema,\label{D2S2}
\ee
it can be decomposed (this is not unique) as
\be
N(1)=\bma{cc} 1 & a \\ . & 1 \ema,\,
M(1)=\bma{cc} 1 & . \\ . & 1 \ema,\,
N(2)=\bma{cc} b & c \\ . & b \ema,\,
N(1)=\bma{cc} . & 1 \\ . & . \ema.
\ee
When the number of states is three there is still more freedom and there
are more varied solutions of this type, but we will not include such
solutions in the list of results. An example is
\[
R:=
\bma{ccc|ccc|ccc}
1 & . & 1 & . & a_5 & a_6 & c_9 & a_8 & a_9 \\
. & 1 & . & . & . & d_8a_5 & . & c_9 & d_8a_8 \\
. & . & 1 & . & . & . & . & . & c_9 \\
\hline
. & . & . & 1 & . & 1 & . & d_8a_5 & d_8a_6 \\
. & . & . & . & 1 & . & . & . & d_8^2a_5 \\
. & . & . & . & . & 1 & . & . & . \\
\hline
. & . & . & . & . & . & 1 & . & 1 \\
. & . & . & . & . & . & . & 1 & . \\
. & . & . & . & . & . & . & . & 1
\ema.
\]

\section{Symmetries}
It is well know that the set of equations (\ref{YBE}) is invariant
under the continuous transformation
\be
R\to \kappa(Q\otimes Q) R (Q\otimes Q)^{-1},\label{QSY}
\ee
where $Q$ is any nonsingular $N\times N$ matrix and $\kappa$ a nonzero
number. They are also invariant under the index reflections
\bea
R_{ij}^{kl} & \to &  R_{kl}^{ij},\label{UDR}\\
R_{ij}^{kl} & \to &  R_{ji}^{lk},\label{FBR}
\eea
the first of these is the usual matrix transposition and the second
follows from $R\to PRP$, where $P_{ij}^{kl}=\delta_i^l\delta_j^k$ is
the permutation operator.

For special choices of $Q$ one can obtain other discrete transformations.
In general, if $\sigma$ is a permutation of the set $\{1,\dots,N\}$,
then $Q^i_j=\delta^i_{\sigma(j)}$ yields a relabeling of indices
$R_{ij}^{kl} \to R_{\sigma(i)\sigma(j)}^{\sigma(k)\sigma(l)}$. An
important special case of this is given by $Q^i_j=\delta^i_{N-j+1}$
which, when followed by transposition, yields reflection across the
secondary diagonal.

In the present paper we restrict the $R$-matrix to be an upper
triangular matrix, and then it is natural to use only upper triangular
transformation matrices $Q$, which we will call $U$ from now on.
Under such tranformations the upper triangular nature of $R$ is
preserved. Furthermore, one easily finds that the diagonal blocks
transform as
\be
R_{im}^{km} \to U_i^{i'}R_{i'm}^{k'm}U^{-1}{}_{k'}^k.
\ee
Thus, in order to fix the remaining rotational freedom we will just need
to impose conditions on the diagonal blocks.

Normally one would require the canonical form of an upper triangular
matrix be the Jordan canonical form, but this may require
transformation matrices that are not upper triangular. For this reason
we must relax the definition of what is a canonical form. It turns out
that using just upper triangular transformation matrices one can bring
all other upper triangular matrices into one of the following five
`semi-canonical' forms
\bea
C_1&=&\bma{ccc} a & . & . \\ . & b & . \\ . & . & c \ema,\,
C_2=\bma{ccc} a & b & . \\ . & a & . \\ . & . & c \ema,\,
C_3=\bma{ccc} a & . & b \\ . & c & . \\ . & . & a \ema,\nonumber\\
C_4&=&\bma{ccc} a & . & . \\ . & b & c \\ . & . & b \ema,\,
C_5=\bma{ccc} a & b & . \\ . & a & b \\ . & . & a \ema.\label{RJC}
\eea
This is the basis of our classification scheme.

\section{How the equations were solved}
\subsection{Breakdown into smaller sets}
As usual we must use the available symmetries to divide the problem
into several smaller ones. In particular we want to fix the continuous
symmetries related to upper triangular transformation matrices. The
detailed breakdown is given in the Appendix, here we give just the
general idea.

We start with the top diagonal block and transform it into the
semi-canonical Jordan form (\ref{RJC}). We may therefore assume that
the upper block is of the type $C_i$ in (\ref{RJC}). We work through
the cases in opposite order, $C_5$ first. Using a reflection across
the antidiagonal when necessary we may assume that the lowest diagonal
block cannot be reflected up and transformed into anything that has
already been analyzed. Using the notation that $\sim C_n$ is anything
that can be transformed to $C_n$ using upper triangular matrices, we
can write the first division by giving the upper and lower blocks:
\begin{enumerate}
\item up $C_5$, down anything
\item up $C_4$, down anything but $\sim C_5$,
\item up $C_3$, down $\sim C_3, \mbox{ or } \sim C_4, \mbox{ or } \sim C_1$,
\item up $C_2$, down $\sim C_4, \mbox{ or } \sim C_1$,
\item up $C_1$, down $\sim C_1$.
\end{enumerate}
In the third and fourth cases note that an antidiagonal reflection
takes $C_4$ into $C_2$.

This the starting point. Further complications arise when $c=a$ in
$C_3$ or $a=b=c$ in $C_1$, because then they commute with an upper
triangular transformation matrix having units on the diagonal. In
those cases one can use the remaining freedom to operate on the other
diagonal blocks. In the appendix we have given the more detailed
breakdown.

\subsection{Solving the equations}
For each subcase we still have to solve the 729 equations. Some of
these equations vanish automatically because of the upper triangular
ansatz.  It turns out that there are also many simple equations that
factor to linear factors. Since the present equation solver routines
cannot handle well very large set of equations, we decided to start
the solution process by interactively splitting the process into
branches and subbranches based on the simple factorizable equations.
[All algebraic work was done using REDUCE \cite{Hearn}.] The algorithm was
the following:

\begin{enumerate}
\item Initialize, in particular record the expression that cannot
vanish (Ex: $A\ne0$).
\item Are there simple factorizable equations?
\begin{itemize}
\item[Yes:] Choose a simple equation (Ex. $B(A-1)(C-D)$=0) and find
its solutions. Go to 3.
\item[No:] Go to 4
\end{itemize}
\item Are there any solutions that do not break the nonvanishing condition?
\begin{itemize}
\item[Yes:] Add the first solution as an additional rule to the current branch
and create a new branch (which inherits the current assigments and
nonvanishing condition) for each additonal allowed branch, if any.
With each later branch add the condition that the assignment of the
previous branch cannot hold. Go to 2. (Ex: $\{\{\{B => 0\},A\}, \{\{A =>
1\},AB\}, \{\{C => D\},AB(A - 1)\}\}$)
\item[No:] Go to 5
\end{itemize}
\item Are there equations you want to solve by hand?
\begin{itemize}
\item[Yes:] Do it, but if you have to assume something, create also a branch
where the assumption does not hold. Go to 3.
\item[No:]  Solve the remaining equations using `groesolve' \cite{MMN} and
output the allowed solutions. Go to 5.
\end{itemize}
\item Is this the last branch?
\begin{itemize}
\item[Yes:] End.
\item[No:] Take the next branch. Go to 2,
\end{itemize}
\end{enumerate}

A log of the interactive solution process was saved using the unix
`tee' utility, and checked later.  Altogether the printout extended
well over 1000 pages. In the end we had 552 solutions, but naturally
most of them turned out to be of the simple type of Sec.\
\ref{simple}, or subcases of other solutions.

\section{The nontrivial solutions}
As expected, the problem of classifying the solutions is almost as
time consuming as finding them. We must characterize the solutions by
something that is invariant under the possible transformations. The
method we finally chose was to classify first by the number of
different eigenvalues of the $R$-matrix.  In counting the different
eigenvalues one must be careful, because some solutions only seem to
have the required numbed of eigenvalues. [An example is provided by
$diag(1,\xi_1,\xi_1^2, \xi_2^2,\xi_2,\xi_1\xi_2^2,\xi_1,
\xi_1^2\xi_2^2,
\xi_2)$, $\xi_i^3=1$, which seems to have six different eigenvalues.
Since there are only three different cubic roots of unity there are
actually at most three different eigenvalues, arranged in several
ways.]

First a word about notation. The zeroes are represented by dots, for
better readability, $\xi$ is a cubic root of unity in general, $\eta$
is a cubic root $\ne1$, (i.e. $\xi^3=1,\, \eta{}^2=-1-\eta$),
$\epsilon=\pm1$, other greek symbols are also roots of some polynomial
equation given in the text. Symbols with different subscripts have
independent values. Small latin letters $x,y,z,p,q,k,\dots$ are free
parameters.  Capital letters have special properties, given in text.

We would like to repeat again that the simple solutions of Sec.\
\ref{simple} are not included.  Also, solutions related by the
discussed transformations are not mentioned separately, neither are
those obtained from previous solutions by restricting parameters.

\subsection{9 different eigenvalues}
First we have a solution with the full number of 9 different diagonal
elements.
\[
R_{9.1}=
\bma{ccc|ccc|ccc}
1 & . & . & . & . & . & . & . & . \\
. & p & . & 1 - q\eta ^2 & . & . & . & . & . \\
. & . & p^2 & . & (1- q\eta)x & . & (1-q\eta)(1-q\eta^2) & . & . \\ \hline
. & . & . & qp^{-1}\eta ^2 & . & . & . & . & . \\
. & . & . & . & q\eta & . & -q(1-q\eta^2)/x & . & . \\
. & . & . & . & . & qp & . & -\eta ^2 q(1 - q\eta) & . \\ \hline
. & . & . & . & . & . & q^2p^{-2}\eta & . & . \\
. & . & . & . & . & . & . & q^2p^{-1}\eta ^2 & . \\
. & . & . & . & . & . & . & . & q^2
\ema
\]
Here $x$ is a free parameter that can be scaled to 1, since it must be
nonzero. The $q=\eta p^2$ subcase of this solution was presented in
\cite{Cou}.

\subsection{8 different eigenvalues}
The first solution is the multiparameter version of the standard GL(3)
solution, as generalized in \cite{Schir} and \cite{AKR} to the
following that has maximum of 4 free parameters, $p,\,c,\,q,\,k$.
\[
R_{8.1}=
\bma{ccc|ccc|ccc}
1 & . & . & . & . & . & . & . & . \\ . & p & . & 1-q & . & . & .
& . & . \\ . & . & c & . & . & . & 1-q & . & . \\ \hline. & . & . &
qp^{-1} & . & . & . & . & . \\ . & . & . & . & A & . & . & . & .
\\ . & . & . & . & . & qk^{-1} & . & 1-q & . \\ \hline . & . & . & .
& . & . & qc^{-1} & . & . \\ . & . & . & . & . & . & . & k & . \\
. & . & . & . & . & . & . & . & B
\ema
\]
where $A,B\in\{1,-q\}$.

If one imposes further relations among the diagonal entries there can be other
nonzero matrix elements elsewhere. The first such solution is
\[
R_{8.2}=
\bma{ccc|ccc|ccc}
1 & . & . & . & . & x & . & - xp^2\mu^{-1} & .\\. & p & . & 1 - \mu & .
& . & . & . & .\\. & . & p^{-1} & . & . & . & 1 - \mu & . & .\\ \hline
. & . & . & p^{-1}\mu & . & . & . & . & .\\. & . & . & . & 1 & . & . & . &
.\\ . & . & . & . & . & p^{-2}\mu^2 & . & 1 - \mu & .\\ \hline . & . & . & .
& . & . & p\mu & . & .\\. & . & . & . & . & . & . & p^2\mu^{-1} & .\\.
& . & . & . & . & . & . & . & \mu^{3}\ema
\]
where $\mu$ is a root of $(\mu^3-1)(\mu^2+1)=0$. In fact $R_{8.2}$ has
8 different eigenvalues when $\mu^2=-1$ and 7 when $\mu^3=1$, but
we keep these two branches together.  If $x=0$ we get a subcase of
$R_{8.1}$. Other solutions of this type will appear later.

The following is a rather asymmetric solution:
\[
R_{8.3}=
\bma{ccc|ccc|ccc}
1 & . & . & . & . & x & . & . & .\\. & p & . & 1 - \xi & . & . & . &
.  & .\\. & . & p^{-1} & . & . & . & . & . & .\\ \hline . & . & . &
p^{-1}\xi & . & . & . & . & .\\. & . & . & . & 1 & . & . & . & .\\.
& . & . & . & . & p^{-2}\xi^2 & . & . & .\\ \hline . & . & . & . & .
& . & p\xi^{2} & .  & .\\. & . & . & . & . & . & . & p^2 & .\\. & .
& . & . & . & . & . & . & \xi\ema
\]

\subsection{7 different eigenvalues}
The following solution is related to $R_{9.1}$: if in the diagonal of
$R_{9.1}$ the parameter $p$ is restricted to be a cubic root of unity
we can have more nonzero matrix elements:
\[
R_{7.1}=
\bma{ccc|ccc|ccc}
1 & . & . & . & . & xy & . & xy\xi^2(1 - q\eta ) & . \\
. & \xi & . & 1 - q\eta ^2 & . & . & . & . & y\xi(1 - q\eta ^2) \\
. & . & \xi^2 & .& (1 - q\eta)x & . & (1-q\eta)(1-q\eta^2) & . & . \\ \hline
. & . & . & q\xi^2\eta^2 & . & . & . & . & - yq\eta \\
. & . & . & . & q\eta & . & -q(1-q\eta^2)/x & . & . \\
. & . & . & . & . & q\xi & . &  - \eta^2q(1 - q\eta) & . \\ \hline
. & . & . & . & . & . & q^2\xi\eta & . & . \\
. & . & . & . & . & . & . & q^2\xi^2 \eta ^2 & . \\
. & . & . & . & . & . & . & . & q^2
\ema
\]
Here we have two scalable parameters $x$ and $y$, and they can both be
scaled to 1, because $x\ne0$ and $y=0$ leads just to a special case of
$R_{9.1}$.

A restriction on the diagonal elements of $R_{8.1}$ allows
\[
R_{7.2}=
\bma{ccc|ccc|ccc}
1 & . & . & . & x & . & . & . & . \\ . & \epsilon_1 & . & 1-q & . & .
& .  & .  & . \\ . & . & p & . & . & . & 1-q & . & . \\ \hline . & . & . &
q\epsilon_1 & . & . & . & . & . \\ . & . & . & . & -q & . & . & . & .
\\ . & . & . & . & . & p\epsilon_2 & . & 1-q & . \\
\hline . & . & . & . & . & . & p^{-1}q & . & . \\ . & .
& . & . & . & . & . & p^{-1}q\epsilon_2
& . \\ . & . & .  & . & . & . & . & . & A
\ema
\]
where $A\in\{1,-q\}$.

\subsection{6 different eigenvalues}
The following solution is related to $R_{9.1}$ and $R_{7.1}$
\[
R_{6.1}=
\bma{ccc|ccc|ccc}
1 & . & . & . & . & . & . & . & .\\. & p & . & 1 - q^2 & . & . & . & .
& .\\. & . & p^2 & . & (1 - q^2)x & . & (1+q)(1-q)^2 & . & .\\ \hline
. & . & . & q^2p^{-1} & . & . & . & . & .\\. & . & . & . & q & . &
q( 1 - q^2)/x & . & .\\. & . & . & . & . & p & . & (1 - q^2) & .\\
\hline . & . & . & .  & . & . & q^4p^{-2} & . & .\\ . & . & . & . & .
& . & . & q^2p^{-1} & .\\. & . & . & . & . & . & . & . & 1\ema
\]
The $p=-q$ subcase was found in \cite{AW}.

The next solution is related to $R_{8.2}$: if we have $\mu=1$ in
$R_{8.2}$ there is other freedom available:
\[
R_{6.2}=
\bma{ccc|ccc|ccc}
1 & . & . & . & . & x & . & y & .\\. & p & . & . & . & . & . & . &
.\\. & . & p^{-1} & . & . & . & . & . & .\\ \hline . & . & . & p^{-1}
& . & . & . & . & .\\. & . & . & . & \epsilon & . & . & . & .\\. & . &
. & . & . & p^{-2}\epsilon & . & . & .\\ \hline . & . & . & . & . & .
& p & . & .\\. & . & . & . & . & . & . & p^2\epsilon & .\\. & . & . &
. & . & . & . & . & \epsilon\ema
\]
Here $x$ and $y$ can be scaled to 1, if they are nonzero.

\subsection{5 different eigenvalues}
There were no new nontrivial solutions with 5 different eigenvalues.

\subsection{4 different eigenvalues}

First we have another relative of $R_{8.1}$
\[
R_{4.1}=
\bma{ccc|ccc|ccc}
1 & . & . & . & x & . & . & . & . \\
. & \mu^2 & . & 1-\epsilon i & . & . & . & .
& - i \mu^ 2 y \epsilon \\
. & . & \mu & . & . & . & 1-\epsilon i & . & . \\
\hline
. & . & . & i \mu^2 \epsilon & . & . & . & . & z \\
. & . & . & . & - i \epsilon & . & . & . & . \\
. & . & . & . & . & \mu^3 & . & 1-\epsilon i & . \\
\hline
. & . & . & . & . & . & i \mu^3 \epsilon & . & . \\
. & . & . & . & . & . & . & i \mu \epsilon & . \\
. & . & . & . & . & . & . & . & - i \epsilon\ema
\]
where $\mu^4=1$.

Up to now the diagonal blocks have been of type $C_1$.  The next two
represent a pattern that we will meet several times later on, the
diagonal and (1,3) blocks are of type $C_3$ (more or less) and in
addition there are some nonzero entries on the fourth off-diagonal.

\[
R_{4.2}=
\bma{ccc|ccc|ccc}
1 & . & x & . & . & . & c & . & a \\
. & p & . & . & . & q & . & p (x + c - p y - q) & . \\
. & . & 1 & . & . & . & . & . & c \\
\hline
. & . & . & p^{-1} & . & y & . & . & . \\
. & . & . & . & \epsilon & . & . & . & . \\
. & . & . & . & . & p^{-1} & . & . & . \\
\hline
. & . & . & . & . & . & 1 & . & x \\
. & . & . & . & . & . & . & p & . \\
. & . & . & . & . & . & . & . & 1\ema
\]

\[
R_{4.3}=
\bma{ccc|ccc|ccc}
1 & . & x & . & . & . &  - x & . & -xy \\
. & p & . & . & . & x-y & . & -zp^2 & . \\
. & . & 1 & . & . & . & . & . &  - y \\
\hline
. & . & . & p^{-1} & . & z & . & y-x & . \\
. & . & . & . & \epsilon & . & . & . & . \\
. & . & . & . & . & p^{-1} & . & . & . \\
\hline
. & . & . & . & . & . & 1 & . & y \\
. & . & . & . & . & . & . & p & . \\
. & . & . & . & . & . & . & . & 1\ema
\]

\[
R_{4.4}=
\bma{ccc|ccc|ccc}
1 & . & . & . & x & . & y & . & . \\
. & \epsilon_1 & . & . & . & . & . & y \epsilon_1 & . \\
. & . & 1 & . & . & . & . & . & y \\
\hline
. & . & . & \epsilon_1 & . & . & . & . & . \\
. & . & . & . & \epsilon_2 & . & . & . & . \\
. & . & . & . & . & \epsilon_1 & . & . & . \\
\hline
. & . & . & . & . & . & p & . & . \\
. & . & . & . & . & . & . & \epsilon_1 p & . \\
. & . & . & . & . & . & . & . & p\ema
\]

\subsection{3 different eigenvalues}

The next solution reduces to $R_{8.2}$ if $y=0$, c.f. also $R_{7.1}$
\[
R_{3.1}=
\bma{ccc|ccc|ccc}
1 & . & . & . & . & x & . &  - x \xi_1^2 \xi_2^2 & . \\
. & \xi_1 & . & 1-\xi_2 & . & . & . & . &  - y \xi_1 \\
. & . & \xi_1^2 & . & . & . & 1-\xi_2 & . & . \\
\hline
. & . & . & \xi_1^2 \xi_2 & . & . & . & . & y \\
. & . & . & . & 1 & . & . & . & . \\
. & . & . & . & . & \xi_1 \xi_2^2 & . & 1-\xi_2 & . \\
\hline
. & . & . & . & . & . & \xi_1 \xi_2 & . & . \\
. & . & . & . & . & . & . & \xi_1^2 \xi_2^2 & . \\
. & . & . & . & . & . & . & . & 1\ema
\]

\[
R_{3.2}=
\bma{ccc|ccc|ccc}
1 & . & . & . & . & x & . & y & . \\
. & \xi  & . & . & . & . & . & . & p \\
. & . & \xi^2 & . & . & . & . & . & . \\
\hline
. & . & . & \xi^2 & . & . & . & . & x p \xi y^{-1} \\
. & . & . & . & 1 & . & . & . & . \\
. & . & . & . & . & \xi  & . & . & . \\
\hline
. & . & . & . & . & . & \xi  & . & . \\
. & . & . & . & . & . & . & \xi^2 & . \\
. & . & . & . & . & . & . & . & 1\ema
\]

\[
R_{3.3}=
\bma{ccc|ccc|ccc}
1 & . & . & . & . & x & . & x \xi_1^2 (1 - \xi_2) & . \\
. & \xi_1 & . & . & . & . & . & . & . \\
. & . & \xi_1^2 & . & y & . & . & . & . \\
\hline
. & . & . & \xi_1^2 & . & . & . & . & -\xi_2 x (1-\xi_2)) y^{-1} \\
. & . & . & . & \xi_2 & . & . & . & . \\
. & . & . & . & . & \xi_2^2 \xi_1 & . & -1+\xi_2 & . \\
\hline
. & . & . & . & . & . & \xi_1 & . & . \\
. & . & . & . & . & . & . & \xi_2^2 \xi_1^2 & . \\
. & . & . & . & . & . & . & . & \xi_2\ema
\]

The following reduces to $R_{8.3}$ if $y=0$
\[
R_{3.4}=
\bma{ccc|ccc|ccc}
1 & . & . & . & . & x & . & . & . \\
. & \xi_1 & . & 1 - \xi_2 \xi_1 & . & . & . & . & . \\
. & . & \xi_1^2 & . & y & . & . & . & . \\
\hline
. & . & . & \xi_2 & . & . & . & . & . \\
. & . & . & . & 1 & . & . & . & . \\
. & . & . & . & . & \xi_2^2 & . & . & . \\
\hline
. & . & . & . & . & . & \xi_2^2 & . & . \\
. & . & . & . & . & . & . & \xi_1^2 & . \\
. & . & . & . & . & . & . & . & \xi_2 \xi_1\ema
\]

\subsection{2 different eigenvalues}
Now most of the solution have the previously mentioned structure: the
diagonal and (1,3) blocks are like $C_3$ and in addition there are
some nonzero entries on the fourth off-diagonal.  We list these
solutions in decreasing number of nonzero elements on the
fourth diagonal.

\[
R_{2.1}=
\bma{ccc|ccc|ccc}
1 & . & x & . & p & . & a & . & q\\
. & \epsilon_1 & . & . & . & k & . & a\epsilon_1 + k\epsilon_1
\frac12(\epsilon_2-1) & .\\
. & . & 1 & . & . & . & . & . & a\\
\hline
. & . & . & \epsilon_1 & . & x\epsilon_1 + k\epsilon_1\frac12 (\epsilon_2-1)
& . & k & .\\
. & . & . & . & \epsilon_2 & . & . & . & k^2p^{-1}\\
. & . & . & . & . & \epsilon_1 & . & . & .\\
\hline
. & . & . & . & . & . & 1 & . & x\\
. & . & . & . & . & . & . & \epsilon_1 & .\\
. & . & . & . & . & . & . & . & 1\ema
\]
\[
R_{2.2}=
\bma{ccc|ccc|ccc}
1 & . & y - x & . & p & . &  - y + x & . & \frac12 ( - 2y^2 + x^2)\\
. & \epsilon & . & . & . &  - x & . &  - y\epsilon & .\\
. & . & 1 & . & . & . & . & . &  - (y + x)\\
\hline
. & . & . & \epsilon & . & y\epsilon & . & x & .\\
. & . & . & . & -1 & . & . & . & x^2p^{-1}\\
. & . & . & . & . & \epsilon & . & . & .\\
\hline
. & . & . & . & . & . & 1 & . & y + x\\
. & . & . & . & . & . & . & \epsilon & .\\
. & . & . & . & . & . & . & . & 1\ema
\]

\[
R_{2.3}=
\bma{ccc|ccc|ccc}
1 & . & x & . & a & . &  - x & . &  - xy\\
. & \epsilon & . & . & . &  - y + x & . &  - y\epsilon & .\\
. & . & 1 & . & . & . & . & . &  - y\\
\hline
. & . & . & \epsilon & . & y\epsilon & . & y - x & .\\
. & . & . & . & -1 & . & . & . & .\\
. & . & . & . & . & \epsilon & . & . & .\\
\hline
. & . & . & . & . & . & 1 & . & y\\
. & . & . & . & . & . & . & \epsilon & .\\
. & . & . & . & . & . & . & . & 1\ema
\]

\[
R_{2.4}=
\bma{ccc|ccc|ccc}
1 & . & x & . & p & . & a\epsilon & . & q\\
. & \epsilon & . & . & . & x - k & . & a & .\\
. & . & 1 & . & . & . & . & . & a\epsilon\\
\hline
. & . & . & \epsilon & . & k\epsilon & . & . & .\\
. & . & . & . & -1 & . & . & . & .\\
. & . & . & . & . & \epsilon & . & . & .\\
\hline
. & . & . & . & . & . & 1 & . & x\\
. & . & . & . & . & . & . & \epsilon & .\\
. & . & . & . & . & . & . & . & 1\ema
\]

\[
R_{2.5}=
\bma{ccc|ccc|ccc}
1 & . & x & . & p & . &  - x & . &  - xy\\
. & \epsilon_1 & . & . & . & . & . &  - x\epsilon_1 & .\\
. & . & 1 & . & . & . & . & . &  - y\\
\hline
. & . & . & \epsilon_1 & . & x\epsilon_1 & . & . & .\\
. & . & . & . & \epsilon_2 & . & . & . & .\\
. & . & . & . & . & \epsilon_1 & . & . & .\\
\hline
. & . & . & . & . & . & 1 & . & y\\
. & . & . & . & . & . & . & \epsilon_1 & .\\
. & . & . & . & . & . & . & . & 1\ema
\]

For the following two the diagonal blocks are of type $C_2$. If the
parameter $b$ were zero one could make an index shuffle so that they
would fit into the previously mentioned band form.
\[
R_{2.6}=
\bma{ccc|ccc|ccc}
1 & x & . &  - x &  - x^2 & . & . & . & a\\
. & 1 & . & . &  - x & . & . & . &  - b\\
. & . & \epsilon_1 & . & . &  - x\epsilon_1 & . & . & .\\
\hline
. & . & . & 1 & x & . & . & . & b\\
. & . & . & . & 1 & . & . & . & .\\
. & . & . & . & . & \epsilon_1 & . & . & .\\
\hline
. & . & . & . & . & . & \epsilon_1 & x\epsilon_1 & .\\
. & . & . & . & . & . & . & \epsilon_1 & .\\
. & . & . & . & . & . & . & . & \epsilon_2\ema
\]

\[
R_{2.7}=
\bma{ccc|ccc|ccc}
1 & x & . &  - x & x(x - 2c) & . & . & . &  - bx\\
. & 1 & . & . & x - 2c & . & . & . &  - b\\
. & . & \epsilon_1 & . & . &  - c\epsilon_1 & . & . & .\\
\hline
. & . & . & 1 &  - x + 2c & . & . & . & b\\
. & . & . & . & 1 & . & . & . & .\\
. & . & . & . & . & \epsilon_1 & . & . & .\\
\hline
. & . & . & . & . & . & \epsilon_1 & c\epsilon_1 & .\\
. & . & . & . & . & . & . & \epsilon_1 & .\\
. & . & . & . & . & . & . & . & \epsilon_2\ema
\]

The following are rather sparse, but have more freedom on the diagonal
\[
R_{2.8}=
\bma{ccc|ccc|ccc}
1 & . & . & . & x & a & . & b & c\\
. & \epsilon_1 & . & . & . & . & . & . & .\\
. & . & \epsilon_1 & . & . & . & . & . & .\\
\hline
. & . & . & \epsilon_1 & . & . & . & . & .\\
. & . & . & . & \epsilon_2 & . & . & . & .\\
. & . & . & . & . & \epsilon_2 & . & . & .\\
\hline
. & . & . & . & . & . & \epsilon_1 & . & .\\
. & . & . & . & . & . & . & \epsilon_2 & .\\
. & . & . & . & . & . & . & . & \epsilon_2\ema
\]

\[
R_{2.9}=
\bma{ccc|ccc|ccc}
1 & . & . & . & x & . & . & . & a\\
. & \epsilon_3 & . & . & . & . & . & . & .\\
. & . & \epsilon_1 & . & . & . & . & . & .\\
\hline
. & . & . & \epsilon_3 & . & . & . & . & .\\
. & . & . & . & \epsilon_4 & . & . & . & .\\
. & . & . & . & . & \epsilon_2 & . & . & .\\
\hline
. & . & . & . & . & . & \epsilon_1 & . & .\\
. & . & . & . & . & . & . & \epsilon_2 & .\\
. & . & . & . & . & . & . & . & \epsilon_5\ema
\]

\subsection{One eigenvalue}
The next four solutions have basically the $C_3$ structure, with some
extra elements.
\[
R_{1.1}=
\bma{ccc|ccc|ccc}
1 & . & x & . & . & . &  - x & . &  - xy\\
. & 1 & . & . & . &  - y + x & . &  - x & q\\
. & . & 1 & . & . & . & . & . &  - y\\
\hline
. & . & . & 1 & . & x & . & y - x &  - q\\
. & . & . & . & 1 & . & . & . & .\\
. & . & . & . & . & 1 & . & . & .\\
\hline
. & . & . & . & . & . & 1 & . & y\\
. & . & . & . & . & . & . & 1 & .\\
. & . & . & . & . & . & . & . & 1\ema
\]

\[
R_{1.2}=
\bma{ccc|ccc|ccc}
1 & . & x & . & . & .   & p & . & a\\
. & 1 & . & . & . & p-q & . & q & k\\
. & . & 1 & . & . & .   & . & . & p\\
\hline
. & . & . & 1 & . & x & . & . &  - k\\
. & . & . & . & 1 & . & . & . & .\\
. & . & . & . & . & 1 & . & . & .\\
\hline
. & . & . & . & . & . & 1 & . & x\\
. & . & . & . & . & . & . & 1 & .\\
. & . & . & . & . & . & . & . & 1\ema
\]

\[
R_{1.3}=
\bma{ccc|ccc|ccc}
1 & . & x & . & . & . &  - x & . &  - xy\\
. & 1 & . & . & . & . & . &  - y &  - p\\
. & . & 1 & . & . & . & . & . &  - y\\
\hline
. & . & . & 1 & . & y & . & . & p\\
. & . & . & . & 1 & . & . & . & q\\
. & . & . & . & . & 1 & . & . & .\\
\hline
. & . & . & . & . & . & 1 & . & y\\
. & . & . & . & . & . & . & 1 & .\\
. & . & . & . & . & . & . & . & 1\ema
\]

\[
R_{1.4}=
\bma{ccc|ccc|ccc}
1 & . & x & . & . & . &  - x & . &  - xy\\
. & 1 & . & . & . & . & . &  - 2y + x & p\\
. & . & 1 & . & . & . & . & . &  - y\\
\hline
. & . & . & 1 & . & 2y - x & . & . & .\\
. & . & . & . & 1 & . & . & . & .\\
. & . & . & . & . & 1 & . & . & .\\
\hline
. & . & . & . & . & . & 1 & . & y\\
. & . & . & . & . & . & . & 1 & .\\
. & . & . & . & . & . & . & . & 1\ema
\]

The next one is like $C_2$. The labeling change $2\leftrightarrow 3$ would
make it almost $C_3$ but would not keep the solution upper triangular.
\[
R_{1.5}=
\bma{ccc|ccc|ccc}
1 & x & . &  - x &  - x^2 &  - k + xq & p & k & a\\
. & 1 & . & . &  - x & . & . & q & b\\
. & . & 1 & . & . &  - x & . & . & p\\
\hline
. & . & . & 1 & x & . & . & p-q &  - b\\
. & . & . & . & 1 & . & . & . & .\\
. & . & . & . & . & 1 & . & . & .\\
\hline
. & . & . & . & . & . & 1 & x & .\\
. & . & . & . & . & . & . & 1 & .\\
. & . & . & . & . & . & . & . & 1\ema
\]

The next three solutions have mixed diagonal blocks. It could be
possible to transform the upper block to $C_2$ or $C_3$ but we chose a
different and apparently simpler form defined by the requirement that
the above mentioned relabeling $2\leftrightarrow 3$ keeps the matrix
upper triangular.
\[
R_{1.6}=
\bma{ccc|ccc|ccc}
1 & x & p &  - x &  - xy &  - 2px - k &  - p & k &  - pq\\
. & 1 & . & . &  - y & . & . &  - p & .\\
. & . & 1 & . & . &  - x & . & . &  - q\\
\hline
. & . & . & 1 & y & p & . & . & .\\
. & . & . & . & 1 & . & . & . & .\\
. & . & . & . & . & 1 & . & . & .\\
\hline
. & . & . & . & . & . & 1 & x & q\\
. & . & . & . & . & . & . & 1 & .\\
. & . & . & . & . & . & . & . & 1\ema
\]

\[
R_{1.7}=
\bma{ccc|ccc|ccc}
1 & x & p &  - x &  - xy & c & a &  - c + x( - q + a) & b\\
. & 1 & . & . &  - y &  - q + p & . & a & .\\
. & . & 1 & . & . &  - x & . & . & a\\
\hline
. & . & . & 1 & y & q & . & . & .\\
. & . & . & . & 1 & . & . & . & .\\
. & . & . & . & . & 1 & . & . & .\\
\hline
. & . & . & . & . & . & 1 & x & p\\
. & . & . & . & . & . & . & 1 & .\\
. & . & . & . & . & . & . & . & 1\ema
\]

\[
R_{1.8}=
\bma{ccc|ccc|ccc}
1 & x & q &  - x &  - x^2 &  - k &  - q & k & px\\
. & 1 & . & . &  - x & q & . & . & p\\
. & . & 1 & . & . &  - x & . & . & .\\
\hline
. & . & . & 1 & x & . & . &  - q &  - p\\
. & . & . & . & 1 & . & . & . & .\\
. & . & . & . & . & 1 & . & . & .\\
\hline
. & . & . & . & . & . & 1 & x & .\\
. & . & . & . & . & . & . & 1 & .\\
. & . & . & . & . & . & . & . & 1\ema
\]

Finally we have two solutions whose diagonal blocks are of type $C_5$.
\[
R_{1.9}=
\bma{ccc|ccc|ccc}
1 & 1 & . & -1 & -1 & x & y & y - x & p\\
. & 1 & 1 & . & -1 & -1 & . & y &  - z + y\\
. & . & 1 & . & . & -1 & . & . & y\\
\hline
. & . & . & 1 & 1 & . & -1 & -1 & z\\
. & . & . & . & 1 & 1 & . & -1 & -1\\
. & . & . & . & . & 1 & . & . & -1\\
\hline
. & . & . & . & . & . & 1 & 1 & .\\
. & . & . & . & . & . & . & 1 & 1\\
. & . & . & . & . & . & . & . & 1\ema
\]

In the next one the parametrization is rather complicated and it would be
interesting to understand its origin. [$x=1$ yields a symmetric solution.]
\[
R_{1.10}=
\bma{ccc|ccc|ccc}
1 & 2 & . & -2 &  - 4x & 8x & 4 & . & -4x(7+x)\\
. & 1 & 2 & . &  - 2x & 4 - 6x - 2x^2 & . & 2x(1+x) &  6x(2+x)(-1+x)\\
. & . & 1 & . & . & 2 - 4x & . & . &  - 4x(1-2x)\\
\hline
. & . & . & 1 & 2x &  - 2x(1-x) & -2 &  - 2x(1+x) & -2x(1+3x)(-2+x)\\
. & . & . & . & 1 & 2x & . &  - 2x & 4x(1-2x)\\
. & . & . & . & . & 1 & . & . & 2 - 4x\\
\hline
. & . & . & . & . & . & 1 &  - 2 + 4x & 4(1-2x)(1-x)\\
. & . & . & . & . & . & . & 1 &  - 2 + 4x\\
. & . & . & . & . & . & . & . & 1\ema
\]

\section{Discussion}
In this paper we have given all upper triangular nonsingular solutions
to the constant quantum Yang-Baxter equation (\ref{YBE}), modulo upper
triangular transformations (\ref{QSY}) and reflections
(\ref{UDR},\ref{FBR}) and minus simple solutions of Sec.\ \ref{simple}.

There is not much that we can say now about the solutions, here are
just some random observations.  (i) Some of solutions ($R_{9.1}$,
$R_{8.1}-R_{8.3}$, $R_{7.1}$, $R_{6.1}$, $R_{6.2}$, $R_{4.1}$ and
$R_{3.1}-R_{3.4}$) satisfy also the weight condition $R_{ij}^{kl}=0$,
if $k+l\ne i+j \pmod 3$.  Several others satisfy the same
condition$\pmod 2$, e.g.\ $R_{2.1}-R_{2.5}$.  (ii) The upper triangular
property is of course sensitive to index relabeling, for example
$2\leftrightarrow 3$ often breaks it. Sometimes one can find an upper
triangular transformation after which the solutions stays upper
triangular even after $2\leftrightarrow 3$ exchange.  This happens for
the solutions $R_{8.3},\,R_{4.2},\,R_{4.4}$, $R_{3.4},\,R_{2.4}$,
$R_{2.5}$, $R_{2.9}$, $R_{1.4}$, $R_{1.6}$, $R_{1.7}$ and $R_{1.8}$,
and they are presented in this special form.  (iii) Some solutions show
interesting parametric relations, in particular $R_{9.1}$, $R_{7.1}$,
$R_{6.1}$ and $R_{1.10}$.

Now that we have these constant solutions several natural questions
come up for further study, for example, can one add a spectral
parameter, can one contruct a corresponding universal R-matrix? The
corresponding algebraic structures also need investigation.

\vskip 1cm
\noindent
{\it Note for the readers of the preprint version:} The bibliography
may contain omissions, if you know additional references where
explicit three-state upper triangular solutions have been
presented, please let me know by e-mail to hietarin@utu.fi

\vfill\eject

\vfill\eject
\appendix
\section*{Appendix}
In this appendix we give the breakdown of the original problem to
smaller subsets. The primary classification proceeds by the upper
triangular blocks on the diagonal. The ``number of solutions'' given
below is the number before any trivial solutions or subcases were
eliminated.

\section{Upper block of type $C_5$}
We use scaling freedom to put all nonzero entires in $C_5$ to 1.  The
solution procedure produced 8 solutions, which were immediately
reduced to 3 basic solutions.

\section{Upper block of type $C_4$}
Let us scale so that $b=c=1$ in $C_4$. In principle we could exclude
lower blocks of type $C_5$ from the very beginning, but no such
solutions were found anyway.

\begin{enumerate}
\item $a$ arbitrary, $c_5\ne0$ or $f_7\ne0$:
3 solutions, all have $a=1$.
\item $a\ne1$, $c_5=f_7=0$:
22 solutions
\item $a=1$, $c_5=f_7=0$:
1) $b_7\ne0$,
4 solutions.
2) $b_7=0$, $a_8=1$,
9 solutions.
3) $b_7=0$, $a_8=0$:
35 solutions.
\end{enumerate}

\section{Upper block of type $C_3$}
We can scale $a=b=1$
\subsection{$c_5\ne0$ or $f_7\ne0$}
no solutions

\subsection{$c_5=f_7=0$, $c\ne 1$}
33 solutions

\subsection{$c_5=f_7=0$, $c=1$}
Now $C_3$ commutes with any upper triangular matrix with units on the
diagonal. Using this we can put the lower block also into a
semicanonical form. Solving some of the equations reveals that the
diagonal entries of the lower block are also all $=1$.
If the lower block is of type $C_5$ a reflection takes the system to
one studied before, same holds if the lower block is of type $C_2$. If
the lower block is of type $C_4$, we get one solution.

What remains is a lower block of type $C_3$ with $a=c$, but possibly
with $b=0$. This again commutes with UT transformation matrix with
units on the diagonal, and thus we will use this rotational freedom to
put the center block in a semicanonical form.
\begin{enumerate}
\item Center block of type 5:
no solutions.
\item Center block of type 2:
3 solutions.
\item Center block of type 3:
42 solutions.
\end{enumerate}

\section{Upper block of type $C_2$}
Now lower block must be of type $C_4$ or $C_1$, all others can be
reflected to cases studied before.
\begin{enumerate}
\item $f_7\ne0$:
No solutions.
\item $c_5\ne0$ or $f_7=0$:
2 solutions.
\item $c_3\ne1$, $c_5=f_7=0$:
11 solutions
\item $c_3=1$, $c_5=f_7=0$:
In this case one quickly finds that the lower block must be
\[
\left(\begin{array}{ccc} k & h & g \\ .&k&. \\ .&.&l \end{array}\right).
\]
Here we need to analyse only the case $h=g=0$ due to the following: 1)
If $k=l,\,h\ne0$ we can use using upper triangular transformations to
put $g=0$. Then lower block is of type $C_2$ which reflects to an
upper block of type $C_4$ done earlier. 2) If $k=l,\,h=0,\,g\ne0$
lower block is of type $C_3$, done earlier. 3) If $k\ne l,\,h\ne0$ we
can transform lower block to type $C_2$ without changing the upper
block. 4) In the remaining cases the lower block is diagonal, there were
15 solutions of this type.
\end{enumerate}

\section{Upper block is nonunit diagonal, $c_5\ne0,\,f_7\ne0$}
\label{LE}
Let us normalize the diagonal elements as $(1,b,c)$. Here we may also
assume that the lower block is diagonalizable by an upper triangular
matrix, but maybe not yet in diagonal form. In most cases solving the
first few equations yields directly a diagonal lower block, exceptions
to this are discussed separately.

In general one finds that if $c_5\ne0$ or $f_7\ne0$ then $c_3=b_2^2$,
this implies in particular that in this section we must have
$b_2\ne1$.  After that one also finds quickly that the lower block must
be
\begin{equation}
\left(\begin{array}{ccc} h & . & g \\ .&k&. \\.&.&l \end{array}\right).
\label{lb:1}
\end{equation}

\begin{enumerate}
\item $b^3\ne1,b^2\ne1$:
3 solutions.

\item $b=-1$:
As mentioned above we must now have $c_3=1$.  If now $l=h$ in
(\ref{lb:1}) the system can be reflected to one of the cases studied
before. If $l\ne h$ one can use the transformation
\[
\left(\begin{array}{ccc} 1 & . & y \\ .&1&. \\.&.&1 \end{array}\right),
\]
which commutes with $diag(1,-1,1)$, to put $g=0$.
One finds 4 solutions.
\item $b^3=1,\,b\ne1$:
5 solutions.
\end{enumerate}

\section{Upper block is nonunit diagonal, $c_5\ne0,\,f_7=0$}
The same comments as above holds in this case and we just give the results.
\begin{enumerate}
\item $b^3\ne1,b^2\ne1$:
4 solutions.
\item $b=-1$:
4 solutions.
\item $b^3=1,\,b\ne1$:
6 solutions.
\end{enumerate}

\section{Upper block is nonunit diagonal, $c_5=0,\,f_7\ne0$}
As before, the lower block must be as in (\ref{lb:1}). If $l=h$ we can
reflect the system to case F or C, else we eliminate $g$ and reflect
to case F.

\section{Upper block is nonunit diagonal, $c_5=0,\,f_7=0$}
\subsection{The diagonal elements $(1,b,c)$ are all different}
\label{LH1}
76 solutions

\subsection{$c=1,\,b\ne1$}
Again one finds that the lower block must be of type (\ref{lb:1}) and
as in \ref{LE}.2 one can show that the lower block must in fact be diagonal.
If its diagonal elements are all different we reflect it to \ref{LH1}.  Thus
we get subcases according to the diagonal entries of the lower block:
\begin{enumerate}
\item Lower block $=diag(l,k,l)$:
21 solutions
\item Lower block $=diag(k,k,l)$ or $diag(k,l,l)$:
37 solutions.
\item Lower block $=diag(l,l,l)$:
19 solutions.
\end{enumerate}

\subsection{$b=1,\,c\ne1$}
Now we may assume that the lower block is transformable to a diagonal
form, which has at most two different elements and is not $(h,l,h)$,
all other cases can be reflected to ones studied before.
Then one finds that the lower block must be
\begin{equation}
\left(\begin{array}{ccc} h & g & . \\ .&k&. \\.&.&l \end{array}\right).
\label{lb:2}
\end{equation}
Since the upper block stays invariant under transformations
\[
\left(\begin{array}{ccc} 1 & y & . \\ .&1&. \\.&.&1 \end{array}\right),
\]
we can transform the case $h\ne k$ to a diagonal form while the case
$h=l$ reflects to $C_4$.
\begin{enumerate}
\item Lower block is $diag(h,h,l)$:
36 solutions.
\item Lower block is $diag(h,l,l)$:
18 solutions.
\item Lower block is $diag(l,l,l)$:
36 solutions.
\end{enumerate}

\subsection{$b=c\ne1$}
Again one can trasform the lower block to something that reflects
 to one of the cases before, except for two cases:
\begin{enumerate}
\item Lower block is $diag(h,h,l)$:
17 solutions.
\item Lower block is $diag(l,l,l)$:
40 solutions
\end{enumerate}

\section{Upper and lower blocks are unit matrices}
In this case we have the full trasformation freedom left and we use it
to bring the center block into the semicanonical form (recall $C_2$
reflects to $C_4$)
\begin{enumerate}
\item Center block type $C_5$:
no solutions.
\item Center block type $C_2$:
2 solutions.
\item Center block type $C_3$:
13 solutions.
\item Center block type $C_1$:
Let us call the diagonal elements of the center block as $(d,f,g)$.
1) $d,f,g$ all different,
4 solutions.
2) $g=d$, $f\ne d$,
6 solutions.
3) $d=f$, $f\ne g$,
9 solutions.
4) Center block unit matrix,
9 solutions.
\end{enumerate}

\end{document}